\documentclass{article}

\usepackage{arxiv}

\usepackage[utf8]{inputenc} 
\usepackage[T1]{fontenc}    
\usepackage{hyperref}       
\usepackage{url}            
\usepackage{booktabs}       
\usepackage{amsfonts}       
\usepackage{nicefrac}       
\usepackage{microtype}      
\usepackage{lipsum}
\usepackage{graphicx}
\usepackage{multirow}
\usepackage{bbding}
\usepackage{amsmath}

\graphicspath{ {./images/} }

\title{Revealing the preference for correcting separated aberrations in joint optic-image design}

\author{
 Jingwen Zhou \\
   Zhejiang University\\
  \texttt{22130055@zju.edu.cn} \\
   \And
 Shiqi Chen $*$\\
  Zhejiang University\\
  \texttt{chenshiqi@zju.edu.cn} \\
   \And
 Zheng Ren \\
  Zhejiang University\\
  \And
 Wenguan Zhang \\
  Zhejiang University\\
  \And
 Jiapu Yan \\
  Zhejiang University\\
  \And
 Huajun Feng \\
  Zhejiang University\\
  \And
 Qi Li \\
  Zhejiang University\\
  \And
 Yueting Chen \\
  Zhejiang University\\
  }

\begin{document}
\maketitle
\begin{abstract}
The joint design of the optical system and the downstream algorithm is a challenging and promising task. Due to the demand for balancing the global optimal of imaging systems and the computational cost of physical simulation, existing methods cannot achieve efficient joint design of complex systems such as smartphones and drones. In this work, starting from the perspective of the optical design, we characterize the optics with separated aberrations. Additionally, to bridge the hardware and software without gradients, an image simulation system is presented to reproduce the genuine imaging procedure of lenses with large field-of-views. As for aberration correction, we propose a network to perceive and correct the spatially varying aberrations and validate its superiority over state-of-the-art methods. Comprehensive experiments reveal that the preference for correcting separated aberrations in joint design is as follows: longitudinal chromatic aberration, lateral chromatic aberration, spherical aberration, field curvature, and coma, with astigmatism coming last. Drawing from the preference, a 10\% reduction in the total track length of the consumer-level mobile phone lens module is accomplished. Moreover, this procedure spares more space for manufacturing deviations, realizing extreme-quality enhancement of computational photography. The optimization paradigm provides innovative insight into the practical joint design of sophisticated optical systems and post-processing algorithms.
\end{abstract}


\section{Introduction}
\begin{figure}[t]
\centering
\includegraphics[scale=0.55]{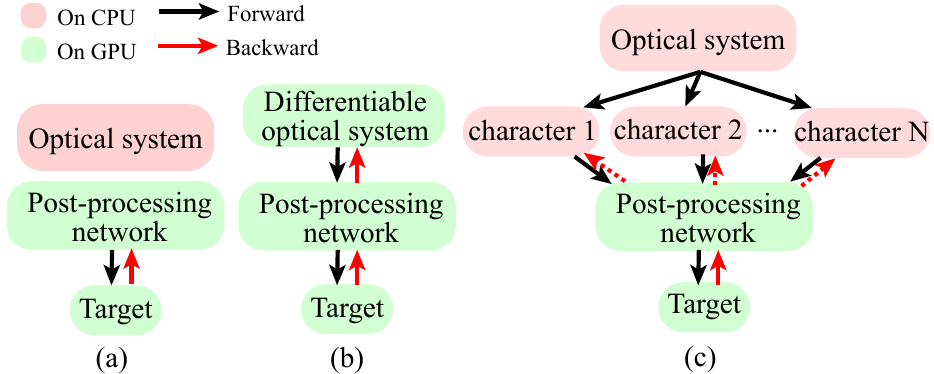}
\caption{\label{introduction}Comparison of the imaging system's design methods: (a) Fixed lens design remedied by post-processing algorithms, (b) End-to-end method, and (c) Ours (Step-by-step joint design).}
\end{figure}

With the popularity of mobile photography (\textit{e.g.}, smartphones, action cameras, drones, \textit{etc.}), the optics of cameras tend to be more miniaturized and lightweight for imaging \cite{zhang2023large, zhang2022end}. However, purely optimizing the hardware configuration is difficult to meet the demand of consumers such as small in size, light in weight, and low in cost. Fortunately, the advancement of \textbf{low-level computer vision} has brought new vitality to the traditional \textbf{optical design and manufacturing}. The engaged representatives can be divided into several categories. One is from \textbf{the perspective of image processing}, the algorithms are designed to remedy the existing defects of the system (\textit{e.g.}, aberration, glare, manufacturing error, \textit{etc.}) \cite{Chung:19, lin2022non, chen2022computational, 10.1145/3474088, 10.1145/2661229.2661260, 10.1145/2516971.2516974, 5753120, zhang2022deep,son2017fast, krishnan2020optical, li2021universal, eboli2022fast}, but they are difficult to influence the front-end lens design, as shown in Fig. \ref{introduction}(a). Another one is \textbf{the consideration of the whole system}, which uses end-to-end strategy to optimize the hardware and the post-processing algorithm in the meanwhile \cite{Li_2022_CVPR, dun2020learned, zhang2022end, baek2021single, ikoma2021depth, lin2022end, yang2023curriculum, sitzmann2018end, sun2020learning, sun2021end, sun2020end, tseng2021neural}, as shown in Fig. \ref{introduction}(b). This manner relies on the differentiable optical propagation to update the system, which means unreasonable computational overhead is required for the optimization of complex systems (please refer to in \textbf{Supplement 1} for detailed analysis). 

After realizing the limitations of existing methods, one may ask that what if we start from \textbf{the perspective of the optical design}? In this box, the lens could be simulated with a comprehensive propagating manner that is relatively close to the real imaging procedure. And besides the traditional optical indicators, \textit{e.g.}, modulation transfer function (MTF), root mean square (RMS), \textit{etc.}, joint optic-image optimization needs to consider the performance of the post-processing system. However, the freedom of optical designing will increase exponentially with the complexity of the system, and the algorithm must realize generalization in various prototypes to prevent the local optimal. Therefore, there are two issues remaining from this perspective, one is how to traverse all possible designs in an orderly manner, another one is how to connect the hardware and the post-processing network when the gradient cannot be propagated.

Compound lens design has various degrees of freedom, including radius, materials, aspheric coefficients, \textit{etc.}, and \textbf{traverse all the potential designs} is time-consuming. Following the principal component analysis in statistics and the vector quantization in auto-regressive, we note that mapping intricate data (\textit{e.g.}, lens configuration) to a high-dimensional representation (\textit{e.g.}, aberrations coefficients) could realize a more comprehensive characterization. To more efficiently traverse the design space anchored by the lens extrinsics, simplicity and completeness of the representation both play important roles. So for the indicators that highly correlated with field-of-view (FoV), \textit{e.g.}, MTF, point diagrams, and point spread function (PSF), the measurement through the whole imaging plane is redundant while the individual of one FoV is insufficient. Moreover, we note that some studies employ Zernike polynomials \cite{conforti1983zernike} to characterize the system, but they have an infinite number of terms theoretically, which couldn't simplify the representation. Fortunately, imaging quality is mostly corrupted by the degradation of primary aberrations, which happens to be quantified by Seidel coefficients \cite{born2013principles}. Therefore, considering the trade-off between accuracy and concision, Seidel is a better option in the view of aberration decomposition. In this way, the design can be traversed in an orderly manner according to the Seidel coefficients. 

As for how to \textbf{connect the optics and the downstream algorithms}, end-to-end methods introduce the differentiable renderers such as Mitsuba2 \cite{nimier2019mitsuba} and Ref. \cite{sun2021end}. While limited by the GPU memory, here the number of rays to be sampled in each pixel is insufficient. When the gradient cannot be propagated, the image simulation system, processing on the CPU, is proposed to bridge the gap between hardware and software. It reproduce the results on sensor plane and then feed them to the post-processing network. In implementation, we note that the simulation system runs by fast fourier transform (FFT) \cite{manual2009optical} or paraxial approximation \cite{born2013principles} is inaccurate for commercial optical designs with large FoVs. To simulate the real imaging procedure, the energy dispersion and the image signal processing (ISP) need comprehensive modeling. Only under these prerequisites, we can minimize the gap between the optics and the downstream algorithms to guide the joint optic-image design. 

This paper aims to \textbf{investigate the preference for correcting separated aberrations in joint optic-image design}. Different from end-to-end methods, we divide the joint design process into three stages and carry out extensive experiments to explore the preference for the post-processing algorithm, so as to redistribute the weights of lens modules and aberration correction methods. In Stage I, a new optics design procedure is developed to separate various samples of aberrations (\textit{e.g.}, spherical aberration, coma, $\cdots$) from a ideal lens prescription. Specifically, we adopt the Seidel coefficients as standards for quantifying different aberrations, and only relax the constraint on one aberration at a time during optimization. Furthermore, to determine the design limits, a comprehensive evaluation experiment is performed. Stage II develops a comprehensive imaging simulation system that characterizes the propagation of rays in the form of complex amplitude when modeling the PSF. After taking the ISP pipeline into account, our imaging simulation system is nearly consistent with real imaging procedure. In Stage III, we propose a network for aberration correction. By introducing FoV information, deformable convolution and multi-scale features, the network is more capable of perceiving the spatially varying aberrations and adapting to the random assembling deviations. The generalization of our proposed network on different aberrations is verified by extensive experiments.

\begin{figure*}[t]
\centering
\includegraphics[scale=0.07]{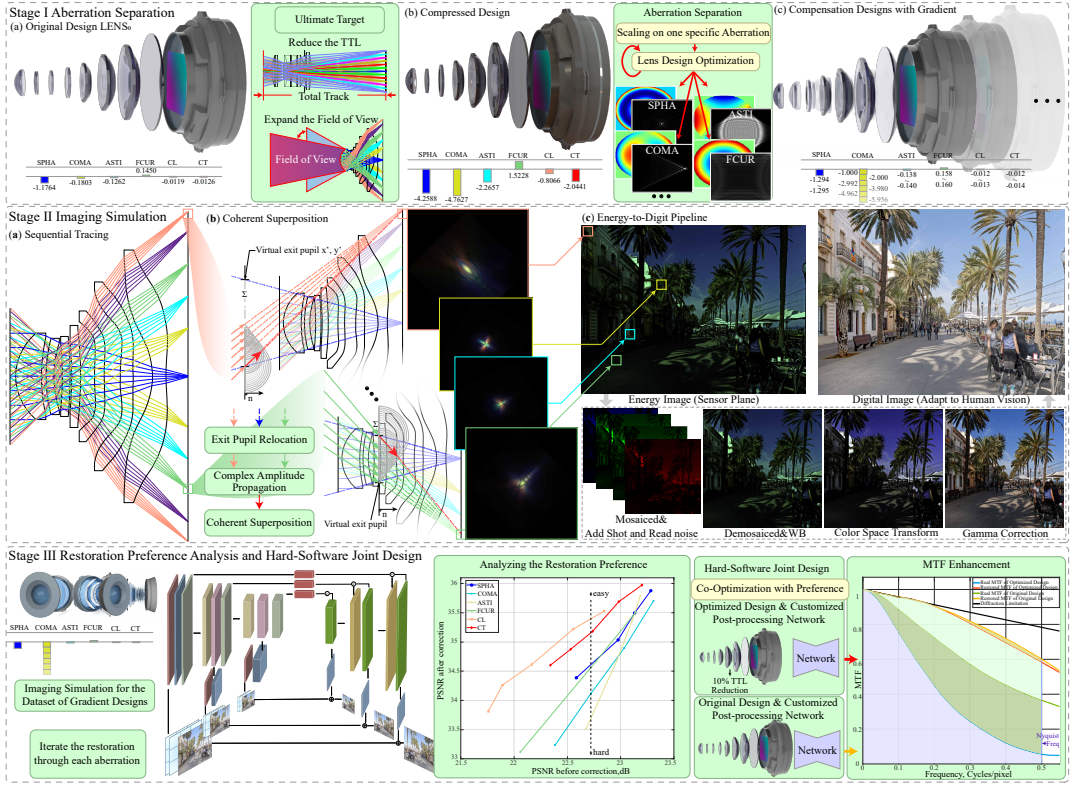}
\caption{\label{1}Overview of our joint optic-image design method.}
\end{figure*}

In a nutshell, our contributions can be summarized as follows:
\begin{itemize}
	\item We reveal the preference for correcting separated aberrations in joint optic-image design (from easy to hard) is: longitudinal chromatic aberration, lateral chromatic aberration, spherical aberration, field curvature, coma, and astigmatism.
 
	\item Considering the priorities of correcting different aberrations, we propose a novel joint design paradigm. Under the guidance of it, the goal of obtaining extreme-quality images with thinner consumer-level mobile phone camera module is achieved.
 
	\item We propose an aberration correction network, which introduces dynamic FoV attention block, deformable residual block, and multi-scale information fusion block. The network can be implemented into the post-processing system to tackle spatially varying aberrations and random manufacturing errors. Extensive experiments demonstrate the superiority of our network over the state-of-the-art methods.
\end{itemize}

The remainder of this paper is organized as follows. Section \ref{Methods} details the method of achieving aberration separation in optic design process and simulating accurate PSFs for synthesizing the aberration-degraded images, together with the illustration of an efficacious aberration correction network. In Section \ref{Experiments and ANALYSIS}, we define the boundaries of optical designs, reveal the preference for correcting separated aberrations, and verify the superiority of our approach through various comparative experiments. Section \ref{application} provides an application example that achieves 10\% total track length (TTL) compression and verifies the robustness of our method for lenses with manufacturing errors. Section \ref{conclusion} concludes this paper.

\section{Methods}
\label{Methods}

Since the enormous complexity of differentiable ray tracing and imaging rendering, existing end-to-end joint design methods are computationally expensive when optimizing complex optical systems with large FoV and multiple high-order aspheric surfaces. Diverse priors are adopted to reduce computational cost and speed up the convergence. Some methods choose to perform the optimization on a well-designed initial structure, which might lead to local optimal. Furthermore, different demands correspond to various initial structures, which greatly reduces the generalization of this strategy. Therefore, to effectively traverse all possible designs in an orderly manner, we characterize the optical system in terms of aberration decomposition and reveal the preference for correcting separated aberrations. Based on this, a novel design paradigm is proposed for joint optic-image design, which can be drawn on by other joint design tasks.

In implementation, we divide the joint design process into three stages. In Section \ref{A}, a new optics design procedure based on Seidel coefficients is developed to separate various samples of aberrations from a ideal lens prescription. Section \ref{B} introduces the imaging simulation system, which allows us to precisely simulate the real imaging procedure and effectively bridge the gap between the optics and the post-processing algorithms. Section \ref{C} presents an aberration correction network. The correction of spatially non-uniform aberrations is realized by introducing dynamic FoV attention blocks, deformable residual blocks, and multi-scale information fusion blocks.

\subsection{Aberration Separation}
\label{A}
In order to summarize the preference for correcting separated aberrations in joint optic-image design, we first design a ideal imaging lens, denoted as $LENS_{0}$. As shown in Table \ref{tab9}, the compressed lenses are designed to have a shorter TTL with a wider FoV, allowing us to optimize them further under the guidance of Seidel coefficients of the ideal lens. There are two common ways to quantify aberrations — Zernike polynomials and Seidel polynomials. Seidel polynomials describe the seven primary aberrations of the optical system while Zernike polynomials can decompose the aberrations more finely with infinite polynomials fitting the deformed wavefront. The reason why we choose Seidel coefficients is that for lightweight optical systems such as mobile lenses, the impact of high-order aberrations is much smaller than that of primary aberrations. Professional designers also prefer the Seidel coefficients to Zernike polynomials in judging whether the lens need to be further optimized. The expressions for seven Seidel coefficients are:

\begin{equation}
\label{eq01}
\left\{\begin{aligned}
S_{I}&=luni\left(i-i^{\prime}\right)\left(i^{\prime}-u\right) \\
S_{I I}&=S_{I} \frac{i_{p}}{i} \\
S_{I I I}&=S_{I I} \frac{i_{p}}{i} \\
S_{I V}&=J^{2} \frac{n^{\prime}-n}{n^{\prime} n r} \\
S_{V}&=\left(S_{I I I}+S_{I V}\right) \frac{i_{p}}{i} \\
C_{I}&=luni\left(\frac{\delta n^{\prime}}{n^{\prime}}-\frac{\delta n}{n}\right) \\
C_{I I}&=C_{I} \frac{i_{p}}{i}
\end{aligned}\right. ,
\end{equation}

\begin{table}[htbp]
\centering
\caption{\bf Extrinsic Parameters of Original and Compressed Lenses}
\begin{tabular}{ccc}
\hline
Extrinsic Parameters & Original & Compressed  \\ \hline
TTL (mm) & 6.0 & 5.4 \\
FoV (°) & 85.6 & 88.1 \\ 
F Number & 1.8 & 1.8 \\
Focal Length (mm) & 5.41 & 5.18 \\  
Optical Distortion & 2.50\% & 2.50\% \\ \hline
\end{tabular}
  \label{tab9}
\end{table}

where $S_{I}$, $S_{II}$, $S_{III}$, $S_{IV}$, $S_V$, $C_I$, and $C_{II}$ respectively represent the primary aberration coefficients of spherical aberration, coma, astigmatism, field curvature, distortion, longitudinal chromatic aberration, and lateral chromatic aberration. $n$ and $n^{'}$ are the refractive indexes of the incident medium and the refraction medium. $u$ and $u^{'}$ are the aperture angles of the object space and the image space. $i$ and $i^{'}$ are the angles between the paraxial rays and the normal of the object space and the image space. $i_p$ is the incident angle of the paraxial chief ray of the whole field-of-view. $l$ is the optical path of the first paraxial ray. $r$ is the radius of the sphere. $J$ is the Lagrange invariant. $\frac{\delta n^{\prime}}{n^{\prime}}$ and $\frac{\delta n}{n}$ are the dispersion of the object space medium and image space medium. It should be emphasized that the aberrations discussed in our work are primary aberrations except for distortion because distortion does not cause blur. The aberrations of the pre-designed ideal lens $LENS_{0}$ are optimized to be as small as possible within the constraints of various design targets. The Seidel coefficients of $LENS_{0}$ are shown in Table \ref{tab1}. $SPHA$, $COMA$, $ASTI$, $FCUR$, $CL$, and $CT$ refer to lenses dominated by spherical aberration, coma, astigmatism, field curvature, longitudinal chromatic aberration, and lateral chromatic aberration respectively. The subscript indicates the index of the lens, and the larger the subscript, the more serious the aberration degradation of the corresponding lens.

\begin{table*}[htbp]
\centering
\caption{\bf The Seidel Coefficients of $LENS_0$ and Compensation Designs with Gradient}
\begin{tabular}{ccccccc}
\hline
  $\space$ & $S_{I}$ & $S_{II}$ & $S_{III}$ & $S_{IV}$ & $C_I$ & $C_{II}$  \\
\hline
$LENS_0$ & -1.176 & -0.180 & -0.126 & 0.145 & -0.012 & -0.012 \\
$SPHA_1$ &  -1.994 & -0.198                         & -0.126                         & 0.147                         & -0.012                         & -0.012                         \\
$SPHA_2$   &  -3.997 & -0.198                         & -0.126                         & 0.145                         & -0.012                         & -0.012                         \\
$SPHA_3$ &  -5.997 & -0.199                         & -0.123                         & 0.144                         & -0.012                         & -0.012                         \\
$SPHA_4$   &  -8.999 & -0.199                         & -0.122                         & 0.142                         & -0.012                         & -0.011                         \\
$SPHA_5$ &  -11.792 & -0.199                         & -0.120                         & 0.142                     & -0.011                         & -0.011                        \\
$COMA_1$ & -1.294                        &  -1.000 & -0.140                         & 0.160                         & -0.012                         & -0.012                          \\
$COMA_2$ & -1.294                         &  -2.000 & -0.140                           & 0.160                         & -0.012                          & -0.012                         \\
$COMA_3$ & -1.294                         &  -2.992 & -0.139                         & 0.159                         & -0.012                         & -0.013                          \\
$COMA_4$ & -1.295                         &  -3.980 & -0.139                         & 0.159                         & -0.013                         & -0.013                        \\
$COMA_5$ & -1.295                         &  -4.962 & -0.138                          & 0.158                         & -0.013                         & -0.013                         \\
$COMA_6$ & -1.295                         &  -5.956 & -0.138                         & 0.158                         & -0.013                         & -0.014                         \\
$ASTI_1$ & -1.165                         & -0.182                         &  -0.251 & 0.155                         & -0.012                         & -0.013                         \\
$ASTI_2$ & -1.165                         & -0.182                         &  -0.503 & 0.160                         & -0.012                         & -0.013                         \\
$ASTI_3$ & -1.165                         & -0.182                         &  -0.755 & 0.155                         & -0.012                         & -0.013                         \\
$ASTI_4$ & -1.159                         & -0.187                         &  -1.008 & 0.142                          & -0.012                         & -0.013                         \\
$ASTI_5$ & -1.141                          & -0.186                         &  -1.259 & 0.144                         & -0.012                         & -0.013                         \\
$FCUR_1$ & -1.188                         & -0.182                         & -0.126                         &  0.201 & -0.012                         & -0.013                         \\
$FCUR_2$ & -1.188                         & -0.182                         & -0.126                         &  0.301 & -0.012                         & -0.013                         \\
$FCUR_3$ & -1.188                          & -0.182                         & -0.126                         &  0.402 & -0.012                         & -0.013                         \\
$FCUR_4$ & -1.188                         & -0.183                         & -0.127                         &  0.503 & -0.012                         & -0.012                         \\
$FCUR_5$   & -1.188                         & -0.183                         & -0.127                         &  0.602 & -0.012                         & -0.012                         \\
$CL_1$ & -1.212                         & -0.185                         & -0.125                         & 0.142                         &  -0.015   & -0.013                         \\
$CL_2$ & -1.212                         & -0.186                         & -0.125                         & 0.141                         &  -0.018  & -0.013                        \\
$CL_3$ & -1.212                         & -0.175                          & -0.128                         & 0.141                         &  -0.022  & -0.013                         \\
$CL_4$ & -1.212                         & -0.175                         & -0.128                         & 0.141                         &  -0.025 & -0.013                         \\
$CL_5$ & -1.212                         & -0.175                         & -0.127                         & 0.141                         &  -0.027 & -0.013                         \\
$CL_6$ & -1.212                         & -0.175                         & -0.127                         & 0.141                         &  -0.030 & -0.013                         \\
$CT_1$ & -1.153                         & -0.175                          & -0.128                         & 0.141                         & -0.012                         &  -0.020    \\
$CT_2$ & -1.163                         & -0.175                          & -0.128                         & 0.141                         & -0.012                         &  -0.025  \\
$CT_3$ & -1.145                         & -0.175                          & -0.128                         & 0.141                         & -0.012                         &  -0.030 \\
$CT_4$ & -1.207                         & -0.175                         & -0.128                         & 0.141                         & -0.012                         &  -0.035 \\
$CT_5$ & -1.212                         & -0.175                         & -0.128                         & 0.141                         & -0.012                         &  -0.040  \\
$CT_6$ & -1.211                         & -0.175                         & -0.128                         & 0.141                         & -0.012                         &  -0.045 \\ \hline
\end{tabular}
  \label{tab1}
\end{table*}

For the compressed lenses, we construct the evaluation function $\phi$ with the Seidel coefficients. $\phi$ can measure the imaging quality of the optical system based on the aberrations that need to be corrected and the parameters which need to be controlled. During optimization, all lens parameters are collectively referred to as variables $x=(x_1,...,x_n )^T$, such as surface parameters and lens thickness; aberrations such as spherical aberration, coma, and astigmatism are defined by Seidel coefficients $S=(S_I, S_{II}, S_{III}, S_{IV}, C_I, C_{II})^T$. There is a very complex nonlinear relationship between the two, which can only be roughly expressed as:
\begin{equation}
\label{eq02}
\left\{\begin{array}{c}
S_I\left(x_{1}, \ldots, x_{n}\right)=S_I \\
\vdots \\
C_{II}\left(x_{1}, \ldots, x_{n}\right)=C_{II}
\end{array}\right.  .
\end{equation}

For the purpose of separating aberrations, the evaluation function $\phi$ of the lenses dominated by different aberrations should be defined separately. For instance, when separating coma, we restrict $S_I, S_{III}, S_{IV}, C_I, C_{II}$ to be consistent with $LENS_0$ while gradually enlarge the value of the coma coefficient $S_{II}$ to $S_{II}^{coma}$, which is sorted from $-1.000$ to $-5.956$ by gradient, as shown in Table \ref{tab1}. The corresponding Seidel coefficients of other optimized designs are also arranged in an orderly manner by gradient, so we refer to these designs as compensation designs with gradient. Due to the fact that the representation of each coefficient is different, the Seidel coefficients are inconsistent with different aberration sensitivities. Under the premise of the same value, different Seidel coefficients bring huge differences in degradation. We can conclude from the data in Table \ref{tab1} and Fig. \ref{2} that the most sensitive coefficients are $C_I$ and $C_{II}$, followed by $S_{II}$, $S_{IV}$, $S_{III}$, and $S_{I}$. The sensitive Seidel coefficients should be sampled more precisely. As an instance, the evaluation function of compensation designs with gradient dominated by coma $\phi_{COMA}$ is defined as:

\begin{equation}
\label{eq03}
\begin{aligned}
\phi_{COMA}=argmin(&|S_I-S_I^0|+ |S_{II}-S_{II}^{coma}|+ |S_{III}-S_{III}^0|+  \\ &|S_{IV}-S_{IV}^0|+ |C_I-C_I^0|+ |C_{II}-C_{II}^0|),
\end{aligned}
\end{equation}

where $S_{I}^0$, $S_{III}^0$, $S_{IV}^0$, $C_I^0$, and $C_{II}^0$ respectively represent the primary aberration coefficients of $LENS_0$, and $S_{II}^{coma} \in\{-1.000,-2.000, -2.992, -3.980, -4.962, -5.956\}$. Considering the conflict between the miscellaneous degradation of lenses and the limited restoring capacity of the aberration correction network, we explore the boundaries of the optical design, which is meticulously elaborated in the experimental section. The sampling ranges and intervals of Seidel coefficients are determined referring to these design boundaries (as shown in Table \ref{tab8}). The corresponding Seidel coefficients of these compensation designs are listed in Table \ref{tab1}, distributed in gradients (highlighted in yellow). Thus, $\phi_{SPHA}$, $\phi_{ASTI}$, $\phi_{FCUR}$, $\phi_{CL}$, and $\phi_{CT}$ can be defined the same way as $\phi_{COMA}$. The damped least squares (DLS) method is applied to find the proper design parameters when $\phi$ approaches its minimum. On the basis of $\phi$, plenty of compensation designs whose degradation are dominated by one kind of aberrations at a time are obtained.

\subsection{Imaging Simulation System}
\label{B}
In our imaging simulation system, the degradation caused by aberrations is characterized through the energy dispersion of the PSFs. Therefore, the accuracy of the PSFs largely determine the authenticity of the simulation images. PSFs obtained under the assumption of Gaussian approximation and FFT are inaccurate when the premise that the image plane is perpendicular to the chief ray is not satisfied. So it is not suitable for imaging systems with large FoV, such as mobile phone lenses. To obtain more accurate PSFs, we refer to ray-tracing and coherent superposition as the basis to build a PSF degradation model.

The surface parameters and materials are defined before performing ray tracing. Apart from  the traditional spherical surface, the most commonly used special surface in mobile phone lens modules is the extended even asphere, which can be defined as
\begin{equation}
z = \frac{cr^{2}}{1+\sqrt{1-(1+k)c^{2}r^{2}}} + \sum_{j=1}^{N} a_jr^{2j},
\label{eq1}
\end{equation}
where $r$ indicates the distance from $(x,y)$ to the z-axis, $r^2=x^2+y^2$, $c$ is the curvature at the apex of the asphere, $k$ is the cone coefficient, and $a_j$ is the coefficient of $r^{2j}$.

Next, the dispersion formulas are introduced to obtain the refractive index of the material at different wavelengths. The Schott formula and The Sellmeier formula are the most common dispersion formulas, defined as:
\begin{equation}
n_{schott}^2(\lambda) = A_0 + A_1\lambda^2 + A_2\lambda^{-2} + A_3\lambda^{-4} + A_4\lambda^{-6} + A_5\lambda^{-8},
\label{eq2}
\end{equation}
\begin{equation}
n_{sellmeier}^2(\lambda) = 1 + \frac{B_1\lambda^2}{\lambda^2-C_1} + \frac{B_2\lambda^2}{\lambda^2-C_2} + \frac{B_3\lambda^2}{\lambda^2-C_3},
\label{eq3}
\end{equation}
where $n$ and $\lambda$ represent the refractive index and the wavelength respectively. $A_0$ to $A_5$, $B_1$ to $B_3$, and $C_1$ to $C_3$ are all dispersion coefficients. After completing the material declaration, the refractive index of the material at specific wavelength can be calculated according to the corresponding formula.

Then, we sample on the entrance pupil. In order to reduce the root mean square error of sampling, we employ Fibonacci sampling, which samples the circular pupil through a helix pattern. The position of the point $(x, y)$ can be expressed as:

\begin{equation}
\label{eq4}
\left\{\begin{array}{l}
x=\sqrt{\frac{j^{\prime}-0.5}{N-0.5}} \cos \left(\frac{4 \pi j^{\prime}}{1+\sqrt{5}}\right) \\
y=\sqrt{\frac{j^{\prime}-0.5}{N-0.5}} \sin \left(\frac{4 \pi j^{\prime}}{1+\sqrt{5}}\right)
\end{array}\right., j^{\prime} \in\{1,2, \ldots, N\},
\end{equation}
where $N$ is the total number of points.

The obtained point $\mathbf{S}=(x,y,z)$ can be regarded as a monochromatic coherent light source, and its propagation direction is determined by the normalized direction vector $\mathbf{D}=(X,Y,Z)$. Thus, the propagation process of light between two surfaces can be defined as:
\begin{equation}
\label{eq5}
\mathbf{S}^{\prime}=\mathbf{S}+t\mathbf{D},
\end{equation}
where $t$ denotes the distance traveled by the ray.

Therefore, the process of ray tracing can be simplified as solving the intersection point $\mathbf{S}^{\prime}$ of the ray and the surface, together with the direction vector $\mathbf{D}^{\prime}$ after refraction. By building the simultaneous equations of Equations (\ref{eq1}) and (\ref{eq5}), the numerical solution $t$ can be acquired through the improved Newton-Raphson method (detailed in \textbf{Supplement 1}). After substituting $t$ into Equation (\ref{eq5}), the intersection point $\mathbf{S}^{\prime}$ can be obtained. According to the Snell's law, the refracted direction vector $\mathbf{D}^{\prime}$ can be computed by:
\begin{equation}
\label{eq6}
\mathbf{D}^{\prime}=\frac{\eta_{1}}{\eta_{2}} \left[\mathbf{D}+\left(\cos \left\langle\mathbf{n}, \mathbf{D}\right\rangle-\sqrt{\frac{\eta_{2}^2}{\eta_{1}^2}-1+\cos^2\left\langle\mathbf{n},\mathbf{D} \right\rangle }\right) \mathbf{n} \right],
\end{equation}
here $\mathbf{n}$ is the normal unit vector of the surface equation, $\eta_1$ and $\eta_2$ are the refractive indices on both sides of the surface, $\mathbf{D}$ is the direction vector of the incident light, and $cos<\cdot,\cdot>$ is the operation for calculating cosine value between two vectors.

By alternately calculating the intersection point $\mathbf{S}^{\prime}$ and the refracted direction vector $\mathbf{D}^{\prime}$, rays can be traced to the exit pupil plane. According to Huygens' principle, the wavefront at the exit pupil plane can be regarded as a collection of secondary wavelet sources. Different from \cite{10.1145/3474088}, we notice that the location and shape of the exit pupils vary depending on the FoV, thus we recalculate the exit pupils for each FoV to appropriately model aberrations. For a certain point $(x_i,y_i,z_i)$ on the image plane, the complex amplitude equals to the coherent superposition of the complex amplitudes of all wavelet sources $(x_e,y_e,z_e)$ propagating from the exit pupil plane to the imaging plane. $\mathbf{\tilde{E}_{x_{i} y_{i}}}$ which can be computed by: 

\begin{equation}
\label{eq7}
\mathbf{\tilde{E}_{x_{i} y_{i}}}=\sum_{y_{e}} \sum_{x_{e}}(\frac{1}{2}a_{0} \left[\cos \left\langle \mathbf{n}, \mathbf{l_{r}}\right\rangle-\cos \left\langle \mathbf{n}, \mathbf{D_{e}}\right\rangle\right]\frac{e^{i k \mathbf{l_{e}}}}{\mathbf{l_{e}}} \frac{e^{i k \mathbf{l_{r}}}}{\mathbf{l_{r}}}),
\end{equation}
where $a_0$ is the amplitude of spherical wave at unit distance. $\mathbf{n}$ and $\mathbf{D_{e}}$ are the normal vector of the exit pupil plane and the direction vector of the ray on the exit pupil plane. $\mathbf{l_{e}}$ and $\mathbf{l_{r}}$ are the total optical path of each ray from the entrance pupil to the exit pupil and from the exit pupil to the image plane, respectively.

Therefore, the intensity $I_{x_{i} y_{i}}$ which represents the PSF matrix can be calculated by:
\begin{equation}
\label{eq9}
I_{x_{i} y_{i}}=\tilde{E}_{x_{i} y_{i}} \cdot * \tilde{E}^{*}_{x_{i}y_{i}},
\end{equation}
where $\tilde{E}^{*}_{x_{i}y_{i}}$ is the complex conjugate of $\tilde{E}_{x_{i} y_{i}}$.

\begin{figure*}[t]
\centering
\includegraphics[scale=0.3]{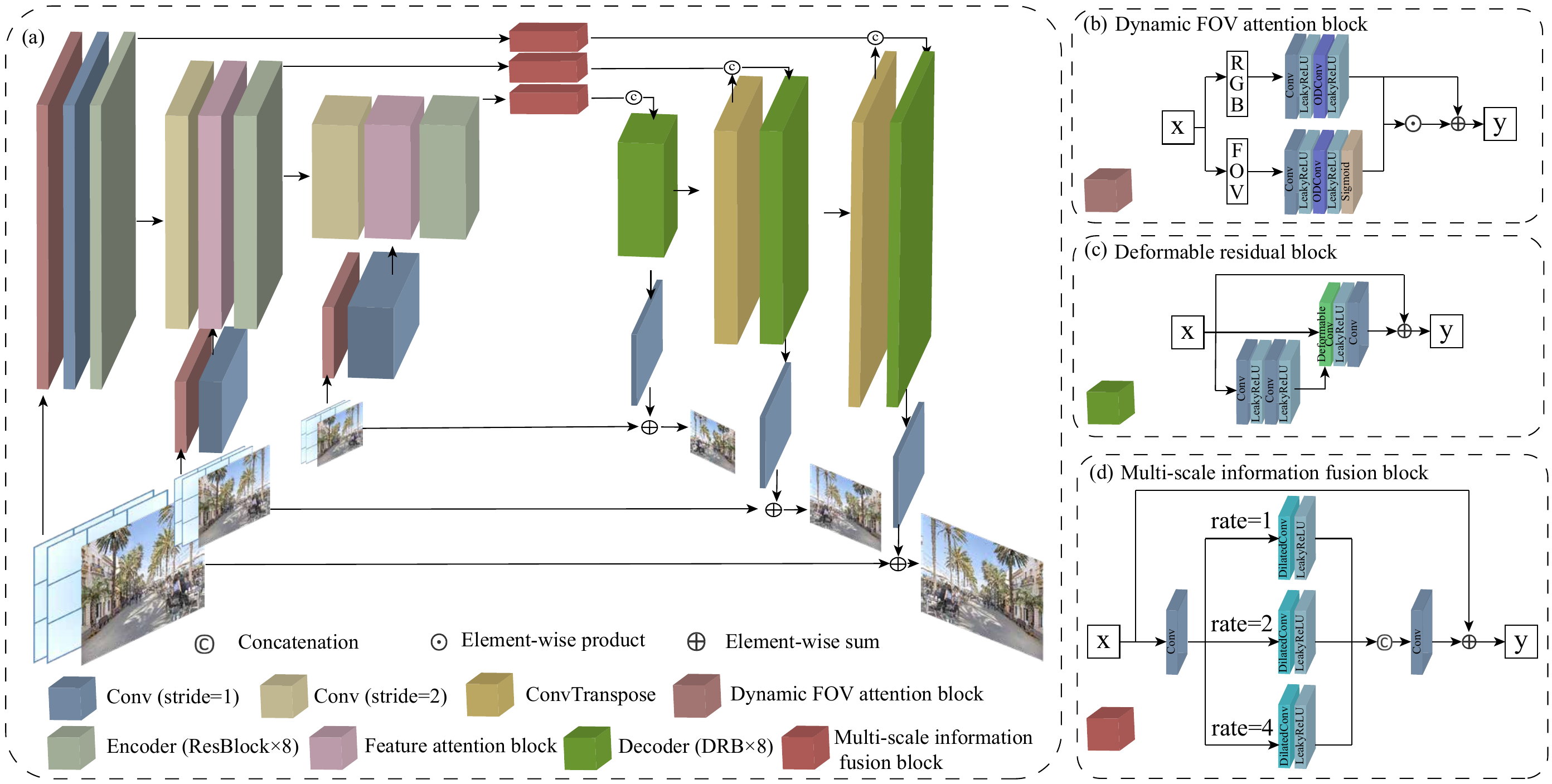}
\caption{\label{7}Aberration correction network architecture. The UNet-based network architecture is shown in (a), and the layer configurations are illustrated with different colored blocks on the bottom. The dynamic FoV attention block, deformable residual block, and multi-scale information fusion block are detailed in (b), (c), and (d) respectively.}
\end{figure*}

After obtaining PSFs of all sampled wavelengths (16 wavelengths varying from 400 nm to 700 nm with a 20-nm interval), we synthesize them into a three-channel RGB PSF based on the spectral sensitivity characteristic of the CMOS as follows:
\begin{equation}
\label{eq15}
PSF_{c}(h, w)=\sum_{\lambda=400}^{700} W_{c}(\lambda) \cdot PSF(h, w, \lambda),
\end{equation}
here, $\lambda$ represents 16 wavelengths varying from 400 nm to 700 nm with a 20-nm interval and $c$ represents R, G, and B channels. $W_{c}(\lambda)$ represents the normalized wavelength response coefficient. $(h, w)$ are the coordinates on the sensor plane which help distinguish the FoV.

As shown in the Stage II of Fig. \ref{7}, a simple ISP pipeline is introduced to help construct realistic aberration-degraded images \cite{Brooks_2019_CVPR}. The energy domain data $I_e$ is partitionally convolved with $PSF_{c}(h, w)$ to synthesize the degraded energy domain patches $D_e(h, w)$. These overlapped patches are center-cropped and concatenated together as $D_e$. Next, we mosaic the degraded raw image $D_e$ before adding shot and read noise to each channel. Moreover, we sequentially apply the demosaic algorithm, white balance (WB), color correction matrix (CCM), and gamma correction (GC) to the R-G-G-B noisy raw image $D_e$, and the aberration-degraded image $I_{d}$ in sRGB domain is obtained. The ISP pipeline can be defined as:
\begin{equation}
\label{eq10}
I_{d}=P_{GC} \circ P_{CCM} \circ P_{WB} \circ P_{demosaic} \circ (P_{mosaic}(D_e)+N),
\end{equation}
where $N$ represents the Gaussian shot and read noise, and $\circ$ is the composition operator. $P_{GC}$, $P_{CCM}$, $P_{WB}$, $P_{demosaic}$, and $P_{mosaic}$ represent the procedures of gamma correction, color correction matrix, white balance, demosaicking, and mosaicking, respectively.

The above steps make up a comprehensive imaging simulation system. In this way, the gap between the optics and the downstream algorithms is eliminated based on this system, which aids in the cooperative optic-image design.

\subsection{Aberration Correction Network}
\label{C}

The image degradation caused by optical aberrations is mainly manifested as blur when we do not consider distortion. Thus, the correction of aberrations can also be regarded as the process of restoring a latent clean image $I_{r}$ from an aberration-degraded blur image $I_{d}$. In this section, we propose an aberration correction network inspired by MIMO-UNet \cite{cho2021rethinking}. The proposed network employs a multiple-input multiple-output framework, and follows a coarse-to-fine correction strategy, enabling the network to handle various aberration deterioration.

\begin{figure*}[ht]
\centering
\includegraphics[scale=0.25]{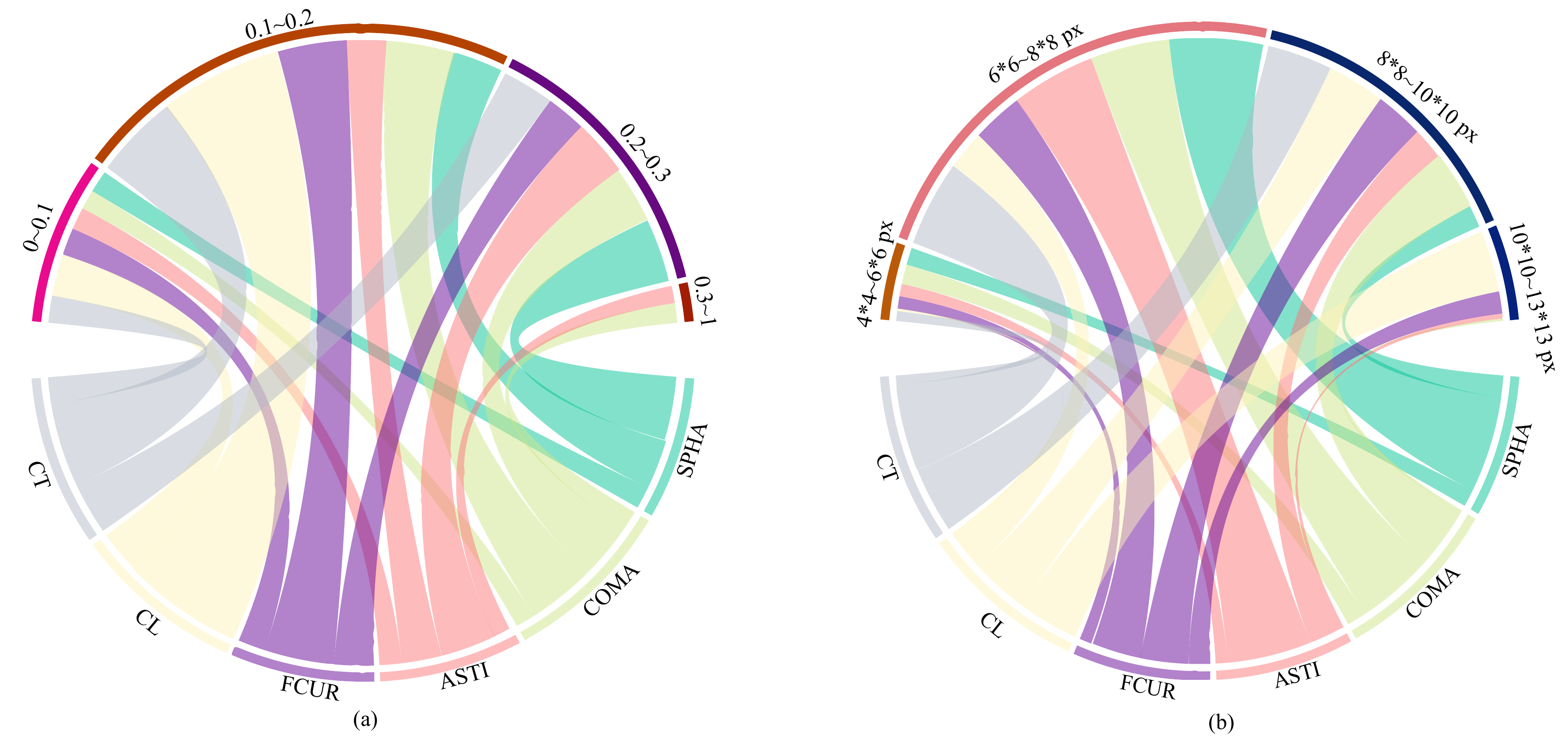}
\caption{\label{9}Distribution analysis of degradation across all designs. From the perspectives of the MTF difference between tangential and sagittal directions (the upper half sector in (a)) and the PSF energy dispersion size (the upper half sector in (b)), (a) and (b) outline the differences in deterioration produced by various types of designs.}
\end{figure*}

The loss function is slightly different from the traditional single-input single-output U-shaped network, that is, the loss function is composed of the sum of loss functions between input and output images of three scales. In order to restore the low-frequency and high-frequency information in the image, we choose L1 loss and FFT loss as the content loss function $L_{cont}$:
\begin{equation}
\label{eq11}
L_{cont}=\sum_{k=1}^{3} \frac{\omega_k}{s_{k}}\left\|I_{r}^{k}-I_{g t}^{k}\right\|_{1}+\sum_{k=1}^{3} \frac{\omega_k}{s_{k}}\left\|\mathcal{F}\left(I_{r}^{k}\right)-\mathcal{F}\left(I_{g t}^{k}\right)\right\|_{1},
\end{equation}
where the subscript $k$ represents the $k^{th}$ layer of the network, $k\in\{1,2,3\}$; $I_r^k$ and $I_{gt}^k$ severally denote the reconstructed images output by the $k^{th}$ layer and the corresponding ground truths; $s_k$ represents the total number of pixels in $I_r^k$, $\omega_k$ represents the weight of the loss function, $\omega_1 = 1, \omega_2 = \omega_3 = 0.8$; $\mathcal{F}(\cdot)$ represents the FFT operation \cite{CHEN_2023_PRL}. At the same time, we define the perceptual loss function $L_{perc}$ based on the pre-trained VGG16 network \cite{simonyan2014very} to ensure that the restored image $I_r$ is more in line with human perception:
\begin{equation}
\label{eq12}
L_{perc}=\sum_{k=1}^{3} \frac{\omega_k}{s_{k}}\left\|\phi_{j}\left(I_{r}^{k}\right)-\phi_{j}\left(I_{g t}^{k}\right)\right\|_{2}^{2},
\end{equation}
here $\phi_{j}(\cdot)$ denotes the feature map obtained by the $j^{th}$ convolutional block in the VGG16. In our work, we choose to use the first block of VGG16 to compute feature maps.

The total loss function of the network can be expressed as follows:
\begin{equation}
\label{eq13}
L_{total}=\lambda_{1} L_{cont}+\lambda_{2} L_{perc},
\end{equation}
where we set $\lambda_{1}=1$, $\lambda_{2}=0.01$.

Unlike common image quality degradation such as motion blur, degradation caused by aberrations has a strong correlation with the FoV. To improve the ability of the network to perceive the spatially non-uniform degradation, we propose a dynamic FoV attention block, which introduces FoV information as a prior. In addition, we replace the static convolution layers in the block with the dynamic ones called ODConv \cite{li2022omni}. ODConv leverages a multi-dimensional attention mechanism with a parallel strategy to learn attentions for convolution kernels along all four dimensions of the kernel space. To be more specific, the parameters of the dynamic FoV attention block can be self-adaptive according to different inputs and the extracted features are modulated by the FoV attention mechanism, which further improves the restoration performance in different FoVs. The block structure is shown in the Fig. \ref{7}(b).

The deformable residual block (DRB) that composes the decoder is an integral part of our proposed network. As shown in Fig. \ref{9} (a), the geometries of degradation caused by different aberrations vary. For example, the tangential and sagittal MTF difference of degradation caused by coma and astigmatism can approach over 0.3 while the others remain within 0.3 (most of them are distributed between 0.1 and 0.2), indicating that the network ought to be highly adaptable to various shapes of aberration degradation. Therefore, the DRB introduces modulated deformable convolution \cite{zhu2019deformable}, which breaks through the inherent geometric structure of the standard convolution operation. The intuitive effect is that the positions and amplitudes of the sampling points in different spatial locations will alter adaptively depending on the image content. Modulated deformable convolution can be defined as:
\begin{equation}
\label{eq14}
y(p)=\sum_{i=-1}^{1} \sum_{j=-1}^{1} w_{i,j} \cdot x\left(p+p_{i,j}+\Delta p_{i,j}\right) \cdot \Delta m_{i,j},
\end{equation}
where $\Delta p_{i,j}$ and $\Delta m_{i, j}$ are the learnable offset and modulation scalar respectively, $\Delta m_{i, j} \in [0, 1]$; $x(p)$, $y(p)$, $p_{i,j}$, and $w_{i,j}$ denote the value at location $p$ from the input feature map, the value at location $p$ of the corresponding output feature map, the relative coordinates within the kernel, and the weight at location $p_{i,j}$ of the kernel. Specifically, we replace the first convolution layer in the ResBlock \cite{he2016deep} with a modulated deformable convolution layer, and the ReLU activation function is likewise substituted by LeakyReLU, as illustrated in Fig. \ref{7}(c). Hence, the ability of the DRB to extract and restore spatially varying features of interest is superior to the common ResBlock.

Empirically, multi-scale features are the key components in aberration correction tasks. U-shaped networks often add skip connections to retain more details, making up for the features damaged by downsampling. This technique, however, can only obtain a single scale of features, which might be ineffective for diverse aberrations with variable sizes and shapes. Fig. \ref{9} (b) shows that the PSF energy dispersion of all designs spans from 4$\times$4 to 13$\times$13, and the associated deteriorations have varied scales. Thus, for the purpose of adapting to multiple scales of aberration deteriorations, we integrate a multi-scale information fusion block (MIFB) between the encoder and decoder at the same level. Each MIFB contains three parallel dilated convolution layers with different dilation rates which are set to 1, 2, and 4 respectively, as shown in Fig. \ref{7}(d). It can perceive multi-scale degradation simultaneously without increasing the parameter amount and computational cost. The features extracted from different receptive fields are then concatenated together and fed to decoders for image reconstruction.

\section{Experiments and Analysis}
\label{Experiments and ANALYSIS}

Because there is no off-the-shelf dataset for our task, we first introduce the procedures of how we simulate the datasets on the basis of the imaging simulation system, together with the implementation details of the training process in Section \ref{AA}. In Section \ref{DD}, we explore the design boundaries of the compressed lenses considering the limited correction capability of the image post-processing algorithm. Section \ref{EE} reveals the preference for correcting separated aberrations after conducting a large number of experiments with 25 qualified lenses. The order of difficulty in correcting separated aberrations is astigmatism > coma > field curvature > spherical aberration > lateral chromatic aberration > longitudinal chromatic aberration. In Section \ref{FF}, we propose a novel joint optic-image design paradigm which redistributes the contributions of the optical module and the post-processing network based on the aberration correction preference. Furthermore, to demonstrate the superiority of our proposed aberration correction network, we compare it with other state-of-the-art methods in Section \ref{CC}. The evaluation results show that our method outperforms these state-of-the-art methods with respect to aberration correction. Meanwhile, our method is validated to be robust to different kinds of aberrations. Section \ref{BB} presents several ablation experiments to verify the practicality of the dynamic FoV attention block, the deformable residual block, and the multi-scale information fusion block.

\subsection{Data Preparation and Implementation Details}
\label{AA}
We adopt DIV8K \cite{gu2019div8k}, which contains 1000 images of 8K resolution as ground truths, and the scenes such as people, animals, natural scenery, \textit{etc.} are covered. Then, we divide these images into the training set, validation set, and test set at 8:1:1. Images of different sizes are center-cropped or resized to 3000$\times$4000 pixels. The degraded images can be obtained by the imaging simulation system proposed in Section \ref{Methods}. First, we assume that the PSFs in the range of 50$\times$50 pixels are spatially uniform, so we only calculate one PSF to represent all the PSFs within 50$\times$50 pixels. After calculating 76800 PSFs (4800 PSFs for each wavelength, and 16 wavelengths varying from 400 nm to 700 nm with a 20-nm interval) for each lens, we synthesize the aberration-degraded images through partitioned convolution operation. The synthetic result of our simulation system and the corresponding digital image are the input and ground truth of our network, respectively. Considering that the imaging process is inevitably affected by noise, we assume that 2$\%$ Gaussian noise is added to the mosaiced input unless otherwise stated. Other data augmentations are carried out including random cropping, random rotating, and shuffle.

During training, the patch size and batch size are set to 128$\times$128 and 64 respectively. The optimizer we use is Adam, with an initial learning rate of 0.0001. After every 50 epochs, the learning rate is reduced to half. The procedure is terminated after 200 epochs. We implement our model in PyTorch \cite{paszke2019pytorch} on a single NVIDIA GeForce RTX 3090 GPU.

\begin{figure}[ht]
\centering
\includegraphics[scale=0.55]{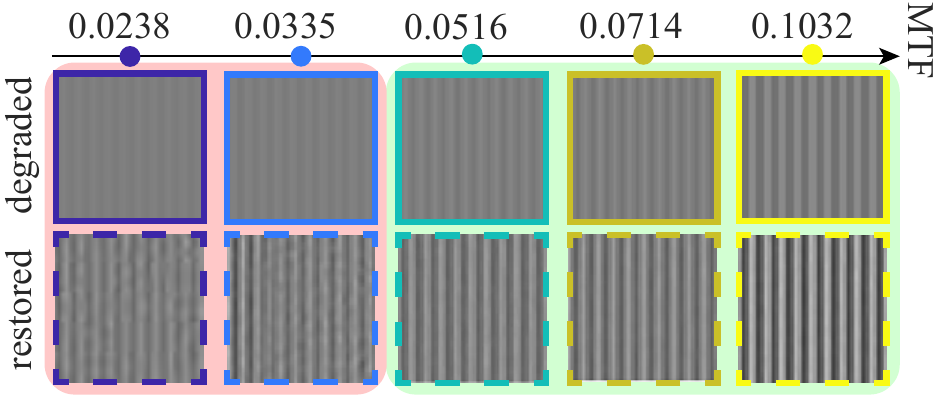}
\caption{\label{8}Degraded line pairs with different MTFs and corresponding restored line pairs. }
\end{figure}

\subsection{Boundaries of the Optical Design}  
\label{DD}
Aberration correction capabilities of the post-processing network affect the boundaries of the optical design. Before revealing the preference for the compensation designs with separated aberrations, we need to explore the restoration limits of the network and determine the boundaries of the optical design. The image restoration results for excessively deteriorated lenses are too poor to be accepted, which has an adverse influence on further analysis of revealing the preference for correcting separated aberrations. Therefore, we need to evaluate the aberration correction effects of different designs and rule out over-degraded ones. For each lens design, we train our proposed network on the  synthetic datasets until it converges. Then, the network is inferred on the corresponding test set and the aberration-corrected images are evaluated.

\begin{figure*}[ht]
\centering
\includegraphics[scale=0.3]{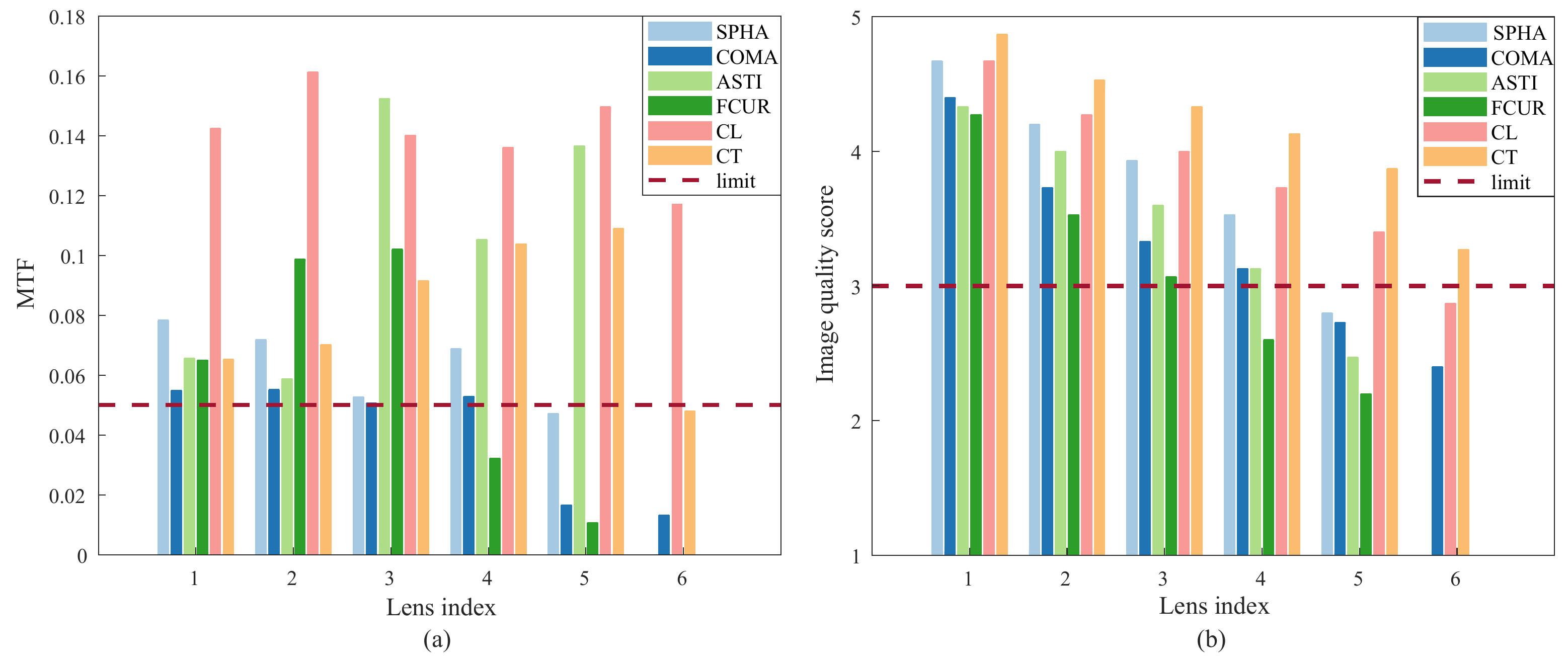}
\caption{\label{3}Evaluation results of all lenses. (a) and (b) represent the evaluation results of images before and after correction, respectively. Bars below the red dot lines demonstrate that the corresponding lenses are considered unqualified.}
\end{figure*}

Artifacts will be introduced into the reconstructed image when the minimum MTF of the degraded image falls below a predetermined threshold. To determine the MTF threshold, we sample a number of line pairs with various MTFs and check to see whether the restored images contain artifacts. As shown in Fig. \ref{8}, with the increase of MTF, artifacts in restored images gradually decrease and disappear when MTF exceeds 0.516. Therefore, to assess the deterioration level of different lenses, we measure the MTF of four key FoVs (0, 0.3, 0.5, and 0.8) of degraded checkerboards and compare the minimum MTF with the threshold. According to the experimental results shown in Fig. \ref{8}, the MTF threshold is set to 0.0516 to prevent introducing artifacts. The evaluation results are shown in Fig. \ref{3}(a). The bars below the red dotted line, which include $SPHA_5$, $COMA_5$, $COMA_6$, $FCUR_4$, $FCUR_5$, and $CT_6$, indicates that the degradation of the corresponding lenses exceeds the correction capability of our aberration correction network. It should be noted that all the restrictions above are for the ideal designs, that is, without considering any manufacturing error.

As for restored images, we mainly assess the image quality from the perspective of human perception, through the image sharpness, color, details, \textit{etc.} Based on the double stimulus impairment scale (DSIS) method \cite{bt2002methodology}, we divide the image quality into 5 levels: excellent, good, fair, poor, and bad, represented by 5, 4, 3, 2, and 1 respectively (visualized in \textbf{Supplement 1}). Excellent means the impairment is imperceptible in the image, and good means the degeneration is perceptible but not annoying. Similarly, fair refers to the image with slightly annoying deterioration while poor refers to the image with annoying damage and bad means the image quality is totally unacceptable. In common sense, the restored image with a score greater than or equal to 3 is considered qualified. We invite 15 observers, some of whom are majored in the field of computer vision while others are nonprofessionals, to rate the reconstructed images. The image to be evaluated and the ground truth are played alternately (the ground truth goes first). Each pair of images are played three times repeatedly, during which the time for observation of each image is 5 seconds. Then a certain time interval (about 3 seconds) is set aside for the observer to rate the restored image. All the scores given by 15 observers are averaged as the subjective evaluation value of the image. As shown in Fig. \ref{3}(b), $SPHA_5$, $COMA_5$, $COMA_6$, $ASTI_5$, $FCUR_4$, $FCUR_5$, and $CL_6$ are not qualified.

Ultimately, we can screen for designs that satisfy all the requirements simultaneously, and they are $SPHA_{1 \sim 4}$, $COMA_{1 \sim 4}$, $ASTI_{1 \sim 4}$, $FCUR_{1 \sim 3}$, $CL_{1 \sim 5}$, and $CT_{1 \sim 5}$. Experiments in the next subsection about revealing preference for correcting separated aberrations are only conducted among these qualified lenses. Furthermore, we can explicitly define the optical design limits with six Seidel coefficients considering the consistency. We compare the Seidel coefficients of these qualified designs in the Table \ref{tab1} and take the maximum value of each coefficient (if coefficients are negative, take the minimum value) as the limit, as shown in the Table \ref{tab8}. These limits can be further applied to optical design stage as prior knowledge to determine sampling ranges and intervals.

\begin{table}[htbp]
\centering
\caption{\bf The Optical Design Limits}
\begin{tabular}{ccccccc}
\hline
  $\space$ & $S_{I}$ & $S_{II}$ & $S_{III}$ & $S_{IV}$ & $C_I$ & $C_{II}$ \\
\hline
limits & -8.999 & -3.980 & -1.008 & 0.402 & -0.027 & -0.040 \\ \hline
\end{tabular}
  \label{tab8}
\end{table}

\subsection{Preference for Correcting Separated Aberrations}
\label{EE}

\begin{figure*}[ht]
\centering
\includegraphics[scale=0.3]{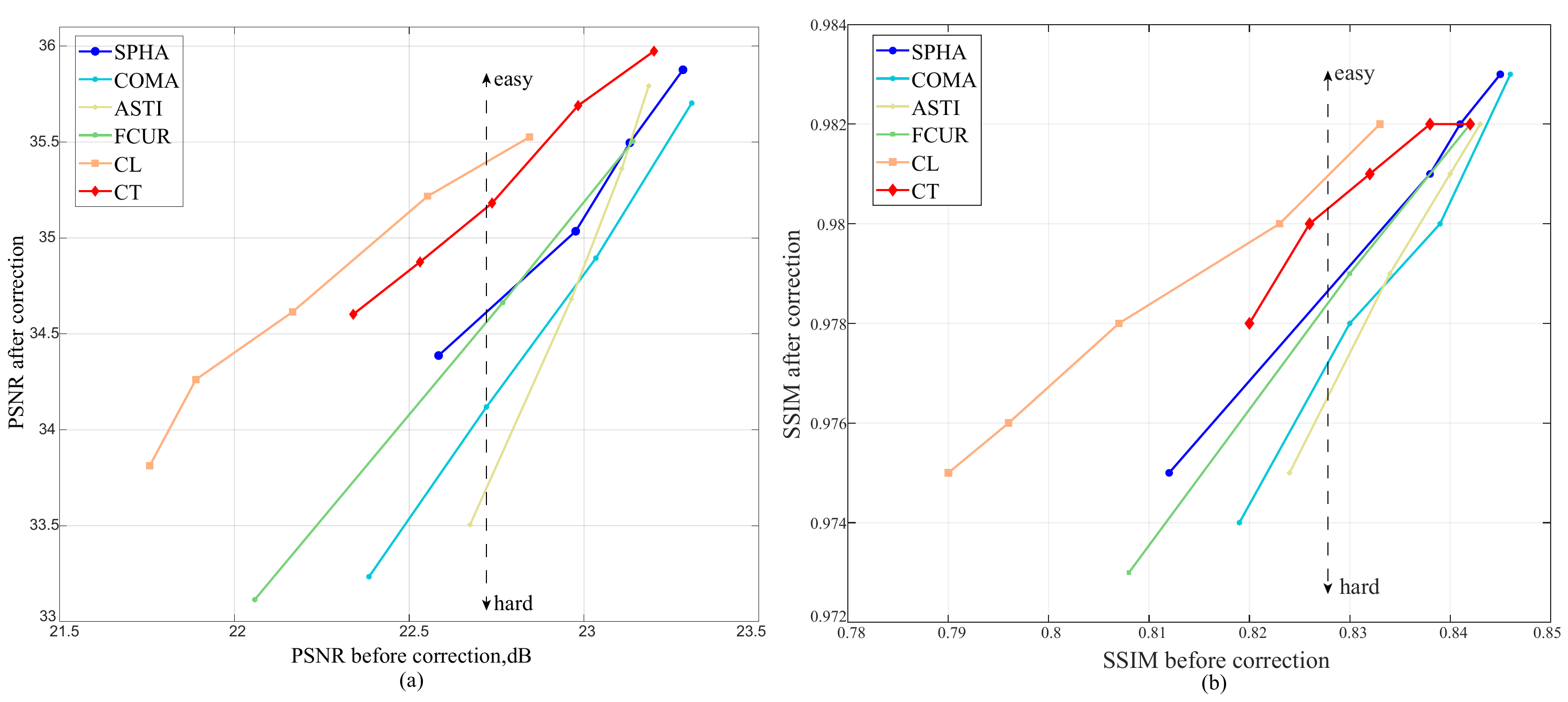}
\caption{\label{2}The line charts of (a) PSNR and (b) SSIM evaluation before \textit{v.s,} after correction.}
\end{figure*}

To reveal the preference for correcting separated aberrations, we need to quantitatively evaluate the performance difference of our aberration correction network on all the qualified lenses mentioned in previous section. Since images in the test set contains various scenes, it is impossible to accurately measure MTF. Therefore, we choose PSNR \cite{johnson2006signal} and SSIM \cite{wang2004image} to quantitatively evaluate the images in the test set before and after correction. Each point in Fig. \ref{2} represents a lens design, and the points of the same color are connected in turn to form a curve that represents the difficulty of correcting this aberration. The reason why we refuse to do direct lens-to-lens comparisons is mainly because of the inconsistency in sensitivity between different Seidel coefficients. It is hard to guarantee that different designs can be degraded to the exact same PSNR or SSIM during the design phase, because there is no clear correspondence between objective evaluation metrics and Seidel coefficients. Fortuitously, comparing the trend of the curves can well avoid the aforementioned problems. Assuming that the PSNR/SSIM of diverse lenses before correction is consistent, the higher position of the curve suggest the better quality of the reconstructed image, which means this kind of aberration is easier to correct for the post-processing algorithm. And the gentler the trend of the curve is, the more adaptive the algorithm is to the aberration. Therefore, it is more challenging for the aberration correction network to rectify the aberrations represented by curves with lower positions and steeper trends.

As shown in Fig. \ref{2}, the orange curve at the top represents the correction results of lenses with longitudinal chromatic aberration $CL_{1 \sim 5}$, and the curve below it represents lateral chromatic lenses $CT_{1 \sim 5}$, followed by designs dominated by spherical aberration $SPHA_{1 \sim 4}$, field curvature $FCUR_{1 \sim 3}$, and coma $COMA_{1 \sim 4}$. The bottom yellow curve refers to astigmatism lenses $ASTI_{1 \sim 4}$. Even if the longitudinal chromatic aberration continues to worsen and the corresponding image quality keeps on deteriorating, the method we propose is still capable of correcting it efficaciously. The average PSNR and SSIM of CL surpass all the other lenses of the same degenerated level. The correction effect of lateral chromatic aberration is slightly worse than that of longitudinal chromatic aberration, but still better than the other four monochromatic aberrations. This is mainly because the input of our aberration correction network contains R, G, and B channels, which naturally leads to strong channel-wise perception. The chromatic aberrations generated by the offset between channels are then aligned effortlessly. Therefore, the cost of correcting chromatic aberrations is the lowest. As for monochromatic aberrations, the performance of correcting spherical aberration surpasses the others due to the fact that spherical aberration is only related to the aperture and is rotationally symmetrical in shape. Coma, astigmatism, and field curvature are related to both aperture and FoV, so they are relatively hard to redress. When there is just minor deterioration, the position of curve $ASTI$ surpasses curve $COMA$ and $FCUR$, even curve $SPHA$. However, we still consider astigmatism as the hardest aberration to be corrected because the the slope of curve $ASTI$ is the steepest, followed by coma and field curvature, as shown in Fig. \ref{2}.

Overall, we can summarize that the order of preference for correcting separated aberrations (from easy to hard) is as follows: longitudinal chromatic aberration, lateral chromatic aberration, spherical aberration, field curvature, and coma, with astigmatism coming last.

\subsection{Joint Optic-image Design Paradigm}
\label{FF}
Based on the preference for correcting separated aberrations concluded in the previous section, the advantage of the post-processing network is correcting longitudinal chromatic aberration, followed by lateral chromatic aberration and spherical aberration. Therefore, more energy can be put into strengthening the constraints on astigmatism, coma and field curvature in order when optimizing the lens.

To further verify the validity of the design paradigm, we compare it with other methods. We first build a simple end-to-end optimization of the initial lens $LENS_0$ on the basis of $dO$ engine \cite{wang2022differentiable}, which achieves the goal of reducing the TTL of the original design without considering the preference for aberration correction. For comparison, we manually optimize the TTL of $LENS_0$ to the same as the end-to-end optimized one. During optimization, two different design strategies are employed. One adopts the conventional lens optimization steps, without considering the correction preference. The other refers to the previous conclusion that the post-processing algorithm has the strongest tolerance for longitudinal chromatic aberration, and therefore we selectively relax the control of it. Finally, we obtain three designs of the same TTL by the end-to-end method, the traditional optical design method and the joint optic-image design method we proposed, which are referred as $LENS_{end2end}$, $LENS_{conventional}$, and $LENS_{ours}$ respectively. Compared with $LENS_0$, the coma, astigmatism and field curvature of $LENS_{end2end}$ and $LENS_{conventional}$ are increased, while the degradation of $LENS_{ours}$ is concentrated in longitudinal chromatic aberration. After the joint optimization, the gap in restoration effects of three lenses can be clearly distinguished from the evaluation results, as shown in Table \ref{tab7}. For the sake of a fair comparison, the degradation levels of the three lenses are almost the same. The lens designed under the guidance of our paradigm obtains the best results of all evaluation metrics, which proves the validity of the aberration correction preference and the effectiveness of our proposed joint design paradigm.

\begin{table}[htbp]
\centering
\caption{\bf Quantitative Evaluation Results of Lenses Optimized by Different Methods Before/After Correction}
\begin{tabular}{cccc}
\hline

  LENS & PSNR$\uparrow$ & SSIM$\uparrow$ & LPIPS \cite{zhang2018unreasonable}$\downarrow$  \\ \hline
$End2end$ & 22.074/32.455 & 0.805/0.971 & 0.305/0.063 \\
$Conventional$ & 22.877/33.953 & 0.828/0.975 & 0.292/0.049 \\ 
$Ours$ & 22.166/\textbf{34.614} & 0.807/\textbf{0.978} & 0.328/\textbf{0.047} \\ \hline
\end{tabular}
  \label{tab7}
\end{table}

\subsection{Recovery Comparison}
\label{CC}

\begin{table*}[htbp]
\centering
\caption{\bf Quantitative Comparison of Different Algorithms on Different Aberrations}
\begin{tabular}{ccccccccccccccccccc}
\hline
                & \multicolumn{3}{c}{SPHA} & \multicolumn{3}{c}{COMA} & \multicolumn{3}{c}{ASTI} \\ \cline{2-10}
                & PSNR$\uparrow$   & SSIM$\uparrow$   & LPIPS$\downarrow$  & PSNR$\uparrow$   & SSIM$\uparrow$   & LPIPS$\downarrow$  & PSNR$\uparrow$   & SSIM$\uparrow$   & LPIPS$\downarrow$  \\ \hline
Degraded images & 21.371 & 0.7502 & 0.3417 & 21.467 & 0.7431 & 0.3927 & 21.968 & 0.7565 & 0.3738 \\
FoV-KPN         & 25.940 & 0.9091 & 0.1449 & 24.760 & 0.8910 & 0.1824 & 26.517 & 0.9135 & 0.1580 \\
MIMO-UNet       & 27.976 & 0.9443 & 0.1080 & 26.512 & 0.9262 & 0.1435 & 28.121 & 0.9405 & 0.1192 \\
MPRNet          & 27.883 & 0.9440 & 0.1012 & 26.051 & 0.9237 & 0.1316 & 28.720 & 0.9473 & 0.1047 \\
Uformer         & 28.180 & 0.9461 & 0.1112 & 26.973 & 0.9327 & 0.1305 & 28.594 & 0.9450	& 0.1145
      \\
Ours   & \textbf{30.592} & \textbf{0.9597} & \textbf{0.0919} & \textbf{29.135} & \textbf{0.9500} & \textbf{0.1100} & \textbf{30.314} & \textbf{0.9546} & \textbf{0.1022} \\ \hline
                &        &        &        &        &        &        &        &        &        \\ \hline
                & \multicolumn{3}{c}{FCUR} & \multicolumn{3}{c}{CL}   & \multicolumn{3}{c}{CT}   \\ \cline{2-10}
                & PSNR$\uparrow$   & SSIM$\uparrow$   & LPIPS$\downarrow$  & PSNR$\uparrow$   & SSIM$\uparrow$   & LPIPS$\downarrow$  & PSNR$\uparrow$   & SSIM$\uparrow$   & LPIPS$\downarrow$  \\ \hline
Degraded images & 20.889 & 0.7222 & 0.4342 & 21.290 & 0.7275 & 0.4408 & 22.060 & 0.7661 & 0.3416 \\
FoV-KPN         & 25.310 & 0.8909 & 0.1963 & 28.371 & 0.9330 & 0.1439 & 28.749 & 0.9404 & 0.1151 \\
MIMO-UNet       & 27.834 & 0.9320 & 0.1391 & 29.603 & 0.9494 & 0.1094 & 30.280 & 0.9586 & 0.0869 \\
MPRNet          & 28.300 & 0.9373 & 0.1271 & 29.944 & 0.9520 & 0.1088 & 31.139 & 0.9634 & 0.0767 \\
Uformer         & 28.140 & 0.9370 & 0.1374 & 30.000 & 0.9530 & 0.1057 & 30.216 & 0.9587	& 0.0896
      \\
Ours  & \textbf{29.593} & \textbf{0.9480} & \textbf{0.1183} & \textbf{31.306} & \textbf{0.9590} & \textbf{0.0964} & \textbf{32.480} & \textbf{0.9680} & \textbf{0.0724} \\ \hline
\end{tabular}
  \label{tab5}
\end{table*}

\begin{table*}[ht]
\centering
\caption{\bf Ablation Study of Different Components}
\begin{tabular}{cccccc}
\hline
Dynamic FoV attention block          & \XSolidBrush & \Checkmark & \Checkmark & \Checkmark & \Checkmark \\
Deformable residual block            & \XSolidBrush & \XSolidBrush & \XSolidBrush & \Checkmark & \Checkmark \\
Multi-scale information fusion block & \XSolidBrush & \XSolidBrush & \Checkmark & \XSolidBrush & \Checkmark  \\ \hline
PSNR                                 & 33.329 & 33.743 & 33.982 & 34.245 & 34.614 \\ \hline
\end{tabular}
  \label{tab4}
\end{table*}

The deterioration caused by aberrations is manifested in the image as spatially varying blur. Considering this premise, we propose a practical blind aberration correction network based on MIMO-UNet. To verify the effectiveness of our proposed method, We choose FoV-KPN \cite{chen2021extreme}, MIMO-UNet, and MPRNet \cite{zamir2021multi} as the representatives of the CNN-based methods. What’s more, we select an emerging methods based on lightweight transformers — Uformer \cite{wang2022uformer}. The above four methods are all blind restoration methods. For a fair comparison, the four methods employ the default setting proposed by their corresponding authors. It should be emphasized that for different aberrations, we design different lenses and generate synthetic datasets based on them. All methods are retrained for different aberrations.

The PSNR, SSIM, and LPIPS indices of these methods on six aberrations are listed in Table \ref{tab5}. Table \ref{tab5} shows that compared with other methods, our proposed network achieves the best performance on PSNR, SSIM, and LPIPS under the circumstances of all aberrations. MIMO-UNet, aiming at solving globally motion blur, cannot obtain extreme-quality performances in correcting optical degradation. Although FoV-KPN also introduces FoV information to help improve the aberration correction capability, the restoration is still not very ideal due to its relatively simple structure. The objective evaluation results of MPRNet are very close to our model, but its parameters and computation overhead greatly surpass our proposed network. Similarly, the computation cost of Uformer is too large due to the existence of self-attention mechanism, compared with simple CNN-based methods. Our proposed method achieving all the optimal results under the circumstances of all aberrations strongly proves that the robustness of our method is better, in the matter of correcting image deterioration caused by optical aberrations.

\begin{figure*}[ht]
\centering
\includegraphics[scale=0.3]{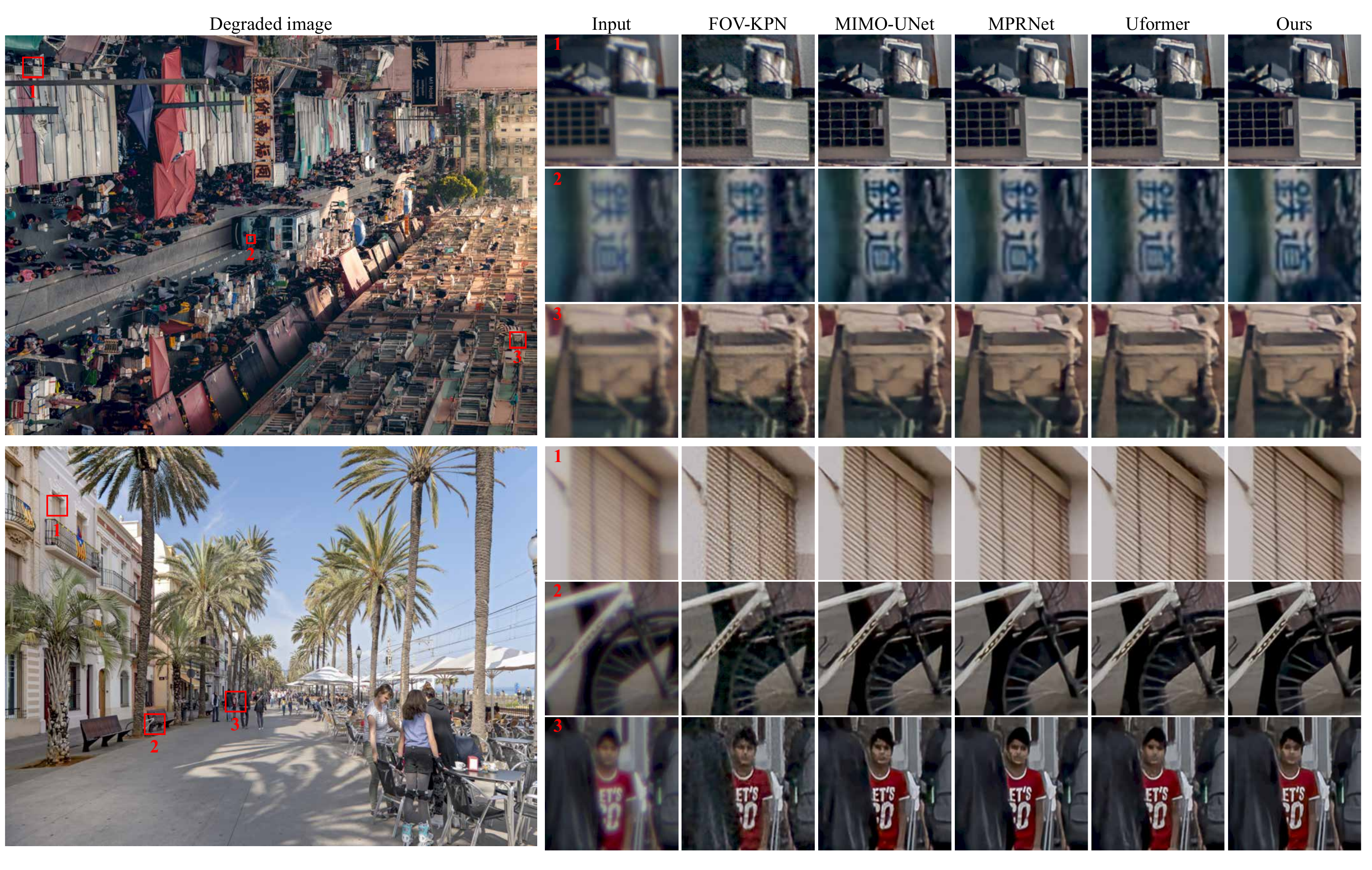}
\caption{\label{6}Aberration correction results with 5$\%$ additive noise. We display full images and several enlarged patches to show fine-grained details.}
\end{figure*}

In Fig. \ref{6}, we visualize the aberration-resolved images for subjective perception. For each image, we show three patches captured from different FoVs, aiming at evaluating the performance of different methods in solving spatially variant aberration blur. The degeneration of the central FoV is imperceptible, and all methods can easily obtain superb performances. However, when it comes to the edge of the FoV, image degradation is much more serious. Methods designed for globally consistent blur struggle to adapt in the face of severe degradation. On the contrary, the proposed model successfully deals with the deterioration with the assistance of dynamic FoV attention blocks and deformable residual blocks, resulting in extreme-quality enhancement. The details of the image restored by our model are still very rich, while other methods tend to lose some textures over the course of correction undesirably. In addition, we add 5$\%$ additive Gaussian noise during data augmentation. Results shown in Fig. \ref{6} demonstrate that our proposed method is relatively more capable of denoising. In brief, our proposed method is capable of sensitively perceiving spatially varying optical aberrations and efficiently integrating multi-scale features for aberration correction.

\begin{figure}[ht]
\centering
\includegraphics[scale=0.22]{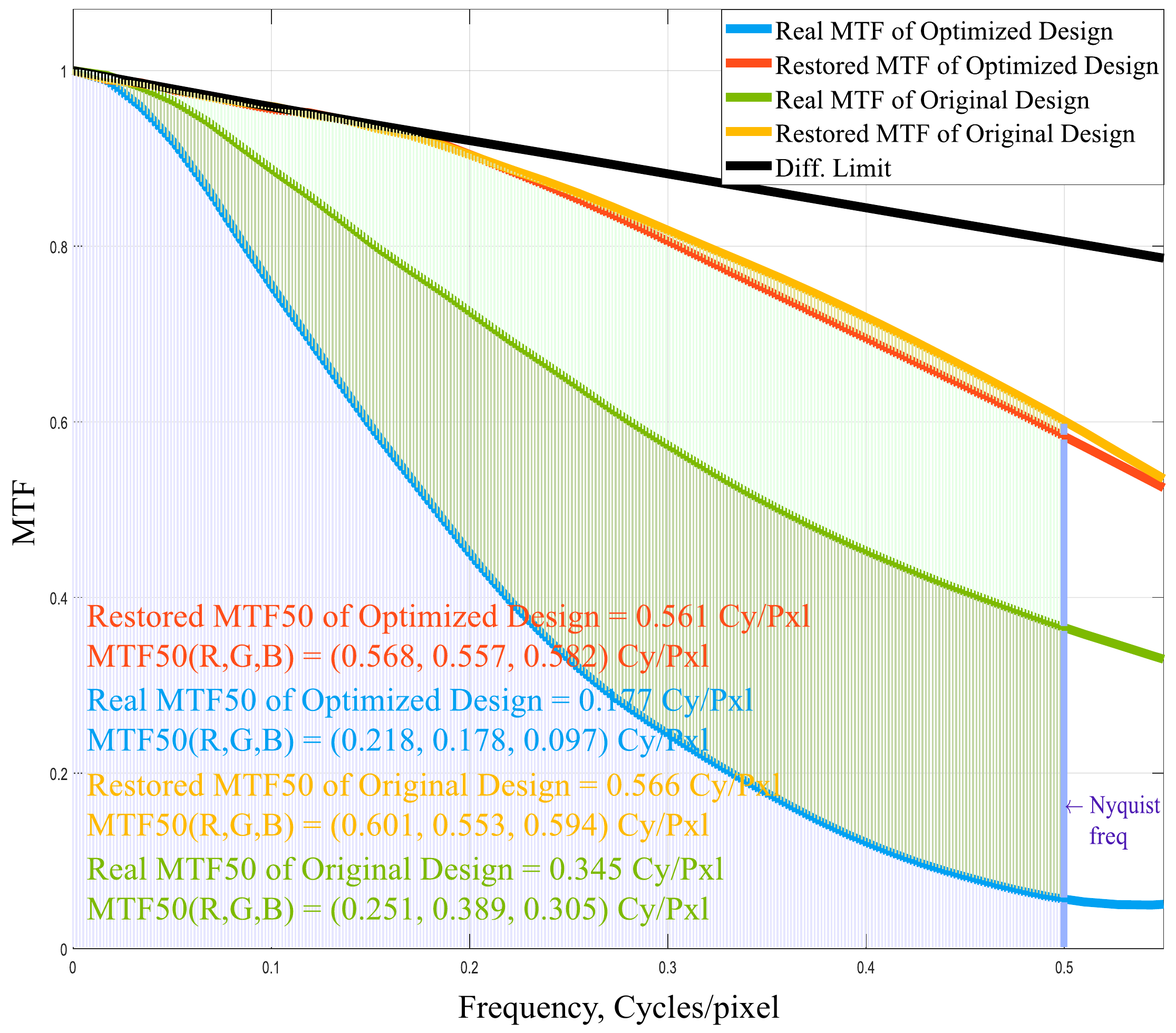}
\caption{\label{4}The MTF enhancement of the original and optimized design. All MTF of the lenses mentioned in this article is calculated by $imatest \textsuperscript{\textregistered}$ for reference.}
\end{figure}

\begin{figure}[ht]
\centering
\includegraphics[scale=0.33]{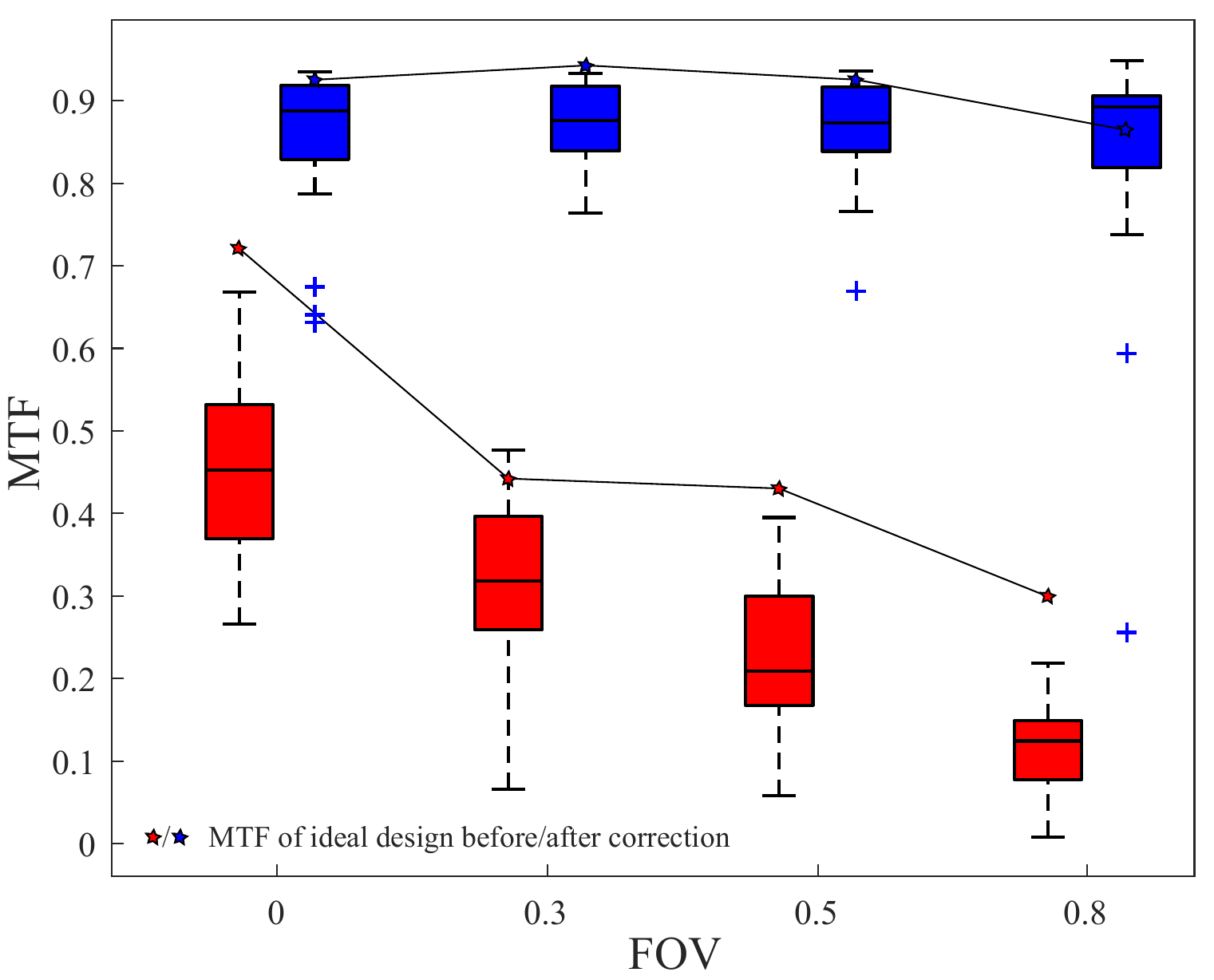}
\caption{\label{5}The MTF enhancement of 20 lenses with tilts and decenters. Red boxes indicates the MTF of the lenses with tilts and decenters before correction in FoV 0, 0.3,0.5, and 0.8, while the blue ones respresents the MTF of the corresponding FoV after correction. The numbers of samples for each box are 20. The red and blue stars indicate the MTF of the ideal design before and after correction. Central black line inside the box: median. Length of the box: interquartile range. Whiskers: maximum and minimum.  Blue Cross: outliers.}
\end{figure}

\subsection{Ablation Study}
\label{BB}
To demonstrate the necessity of three main components in our proposed aberration correction network, we conduct a comprehensive ablation study. To facilitate comparisons, we propose a simplified network. The dynamic FoV attention block is not included in the simplified network. The deformable residual block and the multi-scale information fusion block are replaced by the traditional residual block and the skip connection respectively. The setting of the ablation study is as follows:

\textbf{Ablation on Dynamic FoV Attention Block:} The dynamic FoV attention block adds the FoV information to the feature maps because aberration degradation is strongly correlated with FoV prior. Moreover, the common convolutional layers are replaced with ODConv, which makes the block be more self-adaptive to inputs. The experimental results in the second and third columns show that employing the dynamic FoV attention block greatly improves PSNR by 0.414 dB.

\textbf{Ablation on Deformable Residual Block:} A decoder is constituted by eight deformable residual blocks. Deformable convolution breaks the fixed geometric structures of convolution and adaptively determines the sizes of receptive fields and the weights of different features. The introduction of the deformable residual block increases PSNR from 33.743 dB to 34.245 dB.

\textbf{Ablation on Multi-scale Information Fusion Block:} The existence of the multi-scale information fusion block helps extract multi-scale features, fuse them and communicate encoders and decoders, making the information flow more flexible. Results show that without the multi-scale information fusion block, PSNR drops from 34.614 dB to 34.245 dB.

According to Table \ref{tab4}, we can conclude that the dynamic FoV attention block, the deformable residual block, and the multi-scale information fusion block are indispensable.

\section{Application}
\label{application}

Following the brand-new joint optic-image design paradigm, we \textbf{reduce the 10\% TTL} of the consumer-level mobile phone lens module. We focus on optimizing coma, astigmatism, and field curvature during the optical design stage, leaving the remaining aberrations to the subsequent aberration correction network. The ultimate TTL of the compressed design is 5.4mm. Other information of the lens can be found in \textbf{Supplement 1}. Although the MTF of the compressed design before aberration correction is less than half of the original in high frequency regions, the restoration effect is comparable to it after combining the customized aberration correction network. The MTF of the restored image is greatly enhanced, as shown in Fig. \ref{4}.

In addition, \textbf{manufacturing errors} are inevitably introduced during assembly, resulting in unexpected deviations from the ideal optical design, which is typically manifested as severe and spatially variable degradation in images. Therefore, we need to fine-tune the pre-trained post-processing algorithm to enhance its robustness. Specifically, we randomly sample tilts and decenters as representatives of manufacturing errors in the range of [-0.05, 0.05] and [-0.002, 0.002] respectively and apply them to every surface in the ideal design. Decenters are measured in millimeters, while tilts about the respective axis in a right hand direction are measured in degrees. Among 20 lenses with diverse tolerances, we randomly select 5 of them for fine-tuning. The degradation correction results of all the lenses with tolerance tested on the fine-tuned model is shown in Fig. \ref{5}. It is evident that the fine-tuned model can well adapt to diverse degradation, and reconstruct the imaging results to an acceptable level. Although the MTF of samples with tilts and decenters declines compared to the ideal design after correction (detailed in \textbf{Supplement 1}), the robustness and generalization of our proposed network are greatly improved, mitigating the difficulty in deployment on mass production. Compared with individualized fine-tuning for each lens with random manufacturing deviations, the fine-tuning strategy we propose can greatly reduce the cost and achieve the trade-off between efficiency and effect.

\section{Conclusion}
\label{conclusion}
We explore the preference for correcting separated aberrations in joint design and provide a new optimization paradigm. From the standpoint of the optical design, we characterize the optics with separated aberrations to traverse all possible designs for global optimal. Seidel coefficients are introduced to quantify the aberrations of each design.  Meanwhile, to bridge the front-end lens design and the downstream algorithm in the absence of gradients, an image simulation system is presented, which reproduces the real-world imaging procedure of lenses with large FoVs. Engaging with the image simulation system, we synthetic data pairs, which encode the optical aberrations of corresponding designs, and feed them to the learning-based algorithms for training. In aberration correction stage, we propose a network, which includes FoV information, deformable convolution, and multi-scale features, to perceive and correct the spatially varying aberrations. Extensive experiments demonstrate the superiority of our network over the state-of-the-art methods. Moreover, the preference for correcting separated aberrations in joint design is revealed: longitudinal chromatic aberration, lateral chromatic aberration, spherical aberration, field curvature, coma, and astigmatism. Drawing from the preference, a 10\% reduction in the TTL of the consumer-level mobile phone lens module is accomplished.

Our work strikes a balance between the comprehensiveness of optical designs and the limited computational cost of physical simulation. Additionally, the proposed design paradigm releases more space for manufacturing deviations, realizing extreme-quality enhancement of computational imaging. In conclusion, it is viable to deploy effective joint design of sophisticated optical systems and post-processing algorithms. We expect that this work will stimulate more investigation into joint design methodologies.

\bibliographystyle{unsrt}  





\bibliography{references}

\end{document}